\documentclass{emulateapj}

\def\rpro{\mbox{$r$-process}}
\def\spro{\mbox{$s$-process}}
\def\ncap{\mbox{$n$-capture}}
\def\msun{$M_{\odot}$}

\shorttitle{$r$-process Dispersion in GCs}
\shortauthors{I.~U.\ Roederer}
\submitted{Published in the Astrophysical Journal Letters (2011, 732, L17)}

\begin{document}

\title{Primordial $r$-process Dispersion in Metal-Poor Globular Clusters}

\author{Ian U.\ Roederer}
\affil{Carnegie Observatories,
813 Santa Barbara Street, Pasadena, CA 91101 USA}
\email{iur@obs.carnegiescience.edu}

\begin{abstract}

Heavy elements, those produced by neutron-capture reactions,
have traditionally shown no star-to-star dispersion
in all but a handful of metal-poor globular clusters (GCs).
Recent detections of low [Pb/Eu] ratios or upper limits 
in several metal-poor GCs 
indicate that the heavy elements in these GCs  
were produced exclusively by an $r$-process.
Reexamining GC heavy element abundances from the literature,
we find unmistakable correlations between the [La/Fe] and [Eu/Fe] ratios 
in 4 metal-poor GCs (M5, M15, M92, and \mbox{NGC~3201}),
only 2 of which were known previously.
This indicates that the total 
$r$-process abundances vary star-to-star (by factors of 2--6)
relative to Fe within each GC.
We also identify
potential dispersion in two other GCs (M3 and M13).
Several GCs (M12, M80, and \mbox{NGC~6752}) show no
evidence of $r$-process dispersion.
The $r$-process dispersion is not correlated 
with the well-known light element dispersion,
indicating it was present in the gas throughout the duration
of star formation. 
The observations available at present
suggest that star-to-star $r$-process dispersion within metal-poor
GCs may be a common but not ubiquitous phenomenon that
is neither predicted by nor accounted for 
in current models of GC formation and evolution.

\end{abstract}


\section{Introduction}
\label{intro}

Dispersion among the light elements (C through Al)
is a nearly universal
feature of stars formed in globular cluster (GC) environments.
Heavier $\alpha$ or Fe-group elements (e.g., Ca, Ti, Fe)
do not typically show any star-to-star dispersion within a given GC.
A complex and precisely-tuned---yet common, apparently---sequence 
of events is required to produce these peculiar chemical signatures.
Compared to the lighter elements, however,
the elements produce by neutron ($n$) capture reactions
($Z \gtrsim$~32) are relatively understudied in GCs.

There are two basic mechanisms to
produce nuclei heavier than the Fe-group, reactions that add neutrons
to existing seed nuclei on timescales slow ($s$) or rapid ($r$) relative
to the average $\beta$-decay rates of the radioactive nuclei. 
The main component of the \spro\ is associated with 
low or intermediate mass thermally-pulsing 
asymptotic giant branch (AGB) stars.
Despite more than 2 decades of intense study of field stars
(see the review by \citealt{sneden08}),
the specific astrophysical site(s) of \rpro\ nucleosynthesis is (are)
unknown, 
though association with core-collapse supernovae (SNe)
is likely based on the short timescales required ($\sim$~1~s)
and appearance of \rpro\ material in extremely metal-poor stars
([Fe/H]~$< -$3).\footnote{
For elements X and Y, 
[X/Y]~$\equiv \log_{10} (N_{\rm X}/N_{\rm Y})_{\star} -
\log_{10} (N_{\rm X}/N_{\rm Y})_{\odot}$.}

In metal-poor GCs, \rpro\ nucleosynthesis
dominates production of the heavy elements.
 (e.g., \citealt{gratton04,roederer10b}).
These elements have been observed to
show no significant (i.e., cosmic) star-to-star dispersion within
each GC except for a few rare cases. 
Some massive GCs (e.g., $\omega$~Cen) 
show internal spreads in their heavy elements
(e.g., Ca, Fe, Ba)
and may be the stripped nuclei of
former dwarf galaxies (e.g., \citealt{norris96}).
Individual stars on the AGB in some GCs hint of
self-pollution by \spro\ material \citep{smith08}.
M15 and M92 show an unmistakable star-to-star
dispersion of \rpro\ material relative to Fe
(e.g., \citealt{sneden97,roederer11}),
which is uncorrelated with the light element dispersion
(see also \citealt{dorazi10}).
Although the \rpro\ dispersion in M15 has been known for
more than a decade, it has not been explained.
In this Letter we examine whether M15 and M92 are unique exceptions
among metal-poor GCs with regard to their \rpro\ dispersion.

\section{Literature Data}
\label{data}

We have compiled \ncap\ abundances for individual
GC stars from a number of studies in the literature.
We limit our search to metal-poor GCs ([Fe/H]~$< -$1.0)
whose [La/Fe] and [Eu/Fe] ratios have been derived in a
single study in at least 5~stars.
In practice 10--20~stars are necessary
to reliably identify dispersion in \rpro\ abundances relative to Fe.
Some GCs have been studied by multiple investigators, 
and we examine their abundance correlations independently.
We include data for 11~GCs from 17 
separate studies, listed in Table~\ref{correlatetab}.
Whenever possible, ratios among species of the same ionization
state are given (e.g., [Eu~\textsc{ii}/Fe~\textsc{ii}]).
Typical uncertainties for [La/Fe] and [Eu/Fe] are 0.10--0.15~dex
and not more than 0.20~dex in all cases (except for M92;
see discussion below).
Only two of these GCs, M15 and M92, are previously known to exhibit
star-to-star \rpro\ dispersion. 
M22 and \mbox{NGC~1851} each contain
a population of stars whose heavy elements
were produced only by \rpro\ nucleosynthesis 
and a population with an additional \spro\ component;
we consider only the \rpro\ population.

\begin{deluxetable*}{lccccc}
\tablecaption{Correlations among Abundance Ratios
\label{correlatetab}}
\tablewidth{0pt}
\tabletypesize{\scriptsize}
\tablehead{
\colhead{Cluster (References)} &
\multicolumn{5}{c}{Ratio Pair} \\
\cline{2-6}
\colhead{} &
\colhead{[La/Fe] $+$ [Eu/Fe]} &
\colhead{} &
\colhead{[La/Fe] $+$ [Na/Fe]} &
\colhead{} &
\colhead{[Eu/Fe] $+$ [Na/Fe]}}
\startdata
M3 \citep{sneden04}             & (0.64, 20, 0.00024) & & ($-$0.20, 20, 0.39)  & & (0.12, 22, 0.60)     \\
M3 \citep{cohen05}              & (0.35, 7, 0.45)     & & ($-$0.21, 7, 0.65)   & & ($-$0.24, 8, 0.57)   \\
M5 \citep{ivans01}              & (0.43, 25, 0.034)   & & ($-$0.25, 25, 0.22)  & & ($-$0.16, 25, 0.45)  \\
M5 \citep{lai11}                & (0.80, 17, 0.00010) & & (0.10, 17, 0.70)     & & (0.16, 17, 0.55)     \\
M12 \citep{johnson06}           & (-0.31, 21, 0.17)   & & ($-$0.37, 11, 0.26)  & & (0.41, 11, 0.21)     \\
M13 \citep{sneden04}            & (0.72, 18, 0.00066) & & ($-$0.51, 18, 0.029) & & ($-$0.40, 18, 0.099) \\
M13 \citep{cohen05}             & (0.07, 11, 0.83)    & & ($-$0.23, 12, 0.47)  & & (0.43, 12, 0.16)     \\
M15 \citep{otsuki06}            & (0.97, 6, 0.0011)   & & \nodata              & & \nodata              \\
M15 \citep{sobeck11}            & (0.89, 9, 0.0012)   & & (0.02, 9, 0.95)      & & ($-$0.12, 9, 0.77)   \\
M22 \citep{marino11}\tablenotemark{a} & ($-$0.07, 21, 0.75) & & ($-$0.13, 21, 0.56) & & (0.21, 21, 0.36) \\
M80 \citep{cavallo04}           & (0.37, 8, 0.37)     & & (0.23, 10, 0.52)\tablenotemark{b} & & (0.66, 8, 0.075)\tablenotemark{b} \\
M92 \citep{sneden00b,roederer11} & (0.49, 16, 0.052)  & & ($-$0.02, 13, 0.95)  & & ($-$0.24, 12, 0.45)  \\
NGC~1851 \citep{yong08a}\tablenotemark{a} & (0.50, 5, 0.40) & & (0.41, 5, 0.49) & & (0.52, 5, 0.37)     \\
NGC~1851 \citep{carretta10c}\tablenotemark{a} & (0.87, 7, 0.0079) & & ($-$0.89, 7, 0.0079) & & ($-$0.80, 7, 0.030) \\
NGC~3201 \citep{gonzalez98}     & (0.57, 16, 0.021)   & & ($-$0.24, 16, 0.38)  & & ($-$0.37, 16, 0.16)  \\
NGC~6752 \citep{cavallo04}      & (0.52, 7, 0.23)     & & ($-$0.83, 6, 0.043)\tablenotemark{b} & & ($-$0.71, 8, 0.05)\tablenotemark{b} \\
NGC~6752 \citep{yong05}         & ($-$0.12, 17, 0.66) & & (0.09, 17, 0.74)     & & (0.04, 38, 0.82)     \\
\enddata
\tablenotetext{a}{Using just the \rpro-only stars}
\tablenotetext{b}{[Al/Fe] instead of [Na/Fe]}
\tablecomments{Each set of data indicates $r$, $N$, and $P_{c}(r; N)$.
If two element ratios of a parent distribution are uncorrelated, 
the probability that a random sample of $N$ stars will yield a
correlation coefficient $\geq |r|$ is 
given by $P_{c}(r; N)$.  See, e.g., \citet{bevington03}.
Correlations between [La/Fe] and [Eu/Fe] are significant in
several GCs, while correlations between [La/Fe] or [Eu/Fe] 
and [Na/Fe] are almost never significant.}
\end{deluxetable*}

\section{Globular Clusters with $r$-process Dispersion}
\label{rpro}

In light of new observations of 
\rpro\ abundance patterns in metal-poor field stars, 
we interpret the GC abundance patterns
differently than previous investigators have.
[Pb/Eu] or [Pb/Fe] ratios are more robust 
indicators of \spro\ nucleosynthesis at low metallicity
than [Ba/Eu] or [La/Eu] ratios are
\citep{roederer10b}.
Previously, [Ba/Eu] or [La/Eu] ratios 
enhanced relative to their respective solar \rpro\ ratios
have been interpreted 
as evidence that a small amount of 
\spro\ material is mixed with a more dominant \rpro\ contribution.
In stars with low [Pb/Eu],
we attribute the slightly higher [Ba/Eu] or [La/Eu] to intrinsic 
variations in the \rpro\ abundance patterns themselves.
In M5, M13, M15, M92, and \mbox{NGC~6752}, previous studies
have found [Pb/Eu] ratios or upper limits consistent with
only \rpro\ nucleosynthesis
\citep{yong06,yong08b,sobeck11,roederer11}.
GCs without Pb detections or upper limits have
[La/Eu] ratios similar to those with low [Pb/Eu] ratios,
so it is probable that all of these GCs have
been enriched by \rpro\ but not \spro\ material.

Only three \ncap\ species are routinely studied in metal-poor GCs:
Ba~\textsc{ii}, La~\textsc{ii}, and Eu~\textsc{ii}.
\citet{smith08} has pointed out that GC 
stars on the AGB often have the highest [Ba~\textsc{ii}/Fe] and
[Na~\textsc{i}/Fe] ratios, and he tentatively
attributed this phenomenon to self-enrichment.
Others \citep{shetrone00,ivans01} have noted 
that this may be caused by shortcomings in the analysis
that do not affect La~\textsc{ii} and Eu~\textsc{ii}, so
we focus on 
identifying correlations between [La/Fe] and [Eu/Fe].
If abundances of other heavy elements derived from weak lines 
are available in the literature (e.g., [Nd/Fe]),
we analyze them for confirmation. 
Correlation coefficients and probabilities 
are listed in Table~\ref{correlatetab}.
Six of the GCs (M3, M5, M13, M15, M92, and \mbox{NGC~3201})
show correlated [La/Fe] and [Eu/Fe] with 
less than a 5\% probability that the ratios were drawn from
an uncorrelated parent population.
We now examine each of these GCs in more detail.

\textbf{M15}: 
\rpro\ dispersion in M15 has been
reported by many authors in stars on the red giant branch
\citep{sneden97,sneden00a,sneden00b,otsuki06,sobeck11} and
red horizontal branch \citep{preston06,sobeck11}.
The dispersion spans a range of at least 0.5--0.6~dex in [Eu/Fe], though
no more than 6~stars were studied in any given investigation.
\citet{sneden00b} found a range of nearly 0.9~dex in [Ba/Fe]
in 31~stars.
It is no surprise to find a highly significant correlation
between [La/Fe] and [Eu/Fe] in M15.

\textbf{M92}: 
The correlation between [La/Fe] and [Eu/Fe] is only moderately significant,
but \citet{roederer11} demonstrated that each of these ratios
also correlates strongly with [Ba/Fe] and [Ho/Fe], thus 
strengthening the claim.
That study reported a range of more than 0.8~dex in [Eu/Fe], but
typical uncertainties on each [Eu/Fe] measurement (0.2--0.4~dex)
were significantly
larger than in most studies (0.10--0.15~dex).
\citet{sneden00b} found a range of more than 0.8~dex in [Ba/Fe]
in 32~stars.
We conclude that M92 exhibits significant \rpro\ dispersion.

\textbf{M5}: 
The correlation between [La/Fe] and [Eu/Fe] is significant in
the data of both \citet{ivans01} (25~stars)
and \citet{lai11} (17~stars), and the
Ivans et al.\ correlation becomes even stronger if
one star (\mbox{IV-4}) with low S/N is excluded.
Each of the heavy \ncap\ elements (Ba, La, Ce, Nd, Sm, and Eu) 
studied in at least 10~stars spans a range of 0.25--0.45~dex.
M5 also exhibits a clear signature of 
\rpro\ dispersion.

\begin{figure*}
\begin{center}
\includegraphics[angle=0,width=2.8in]{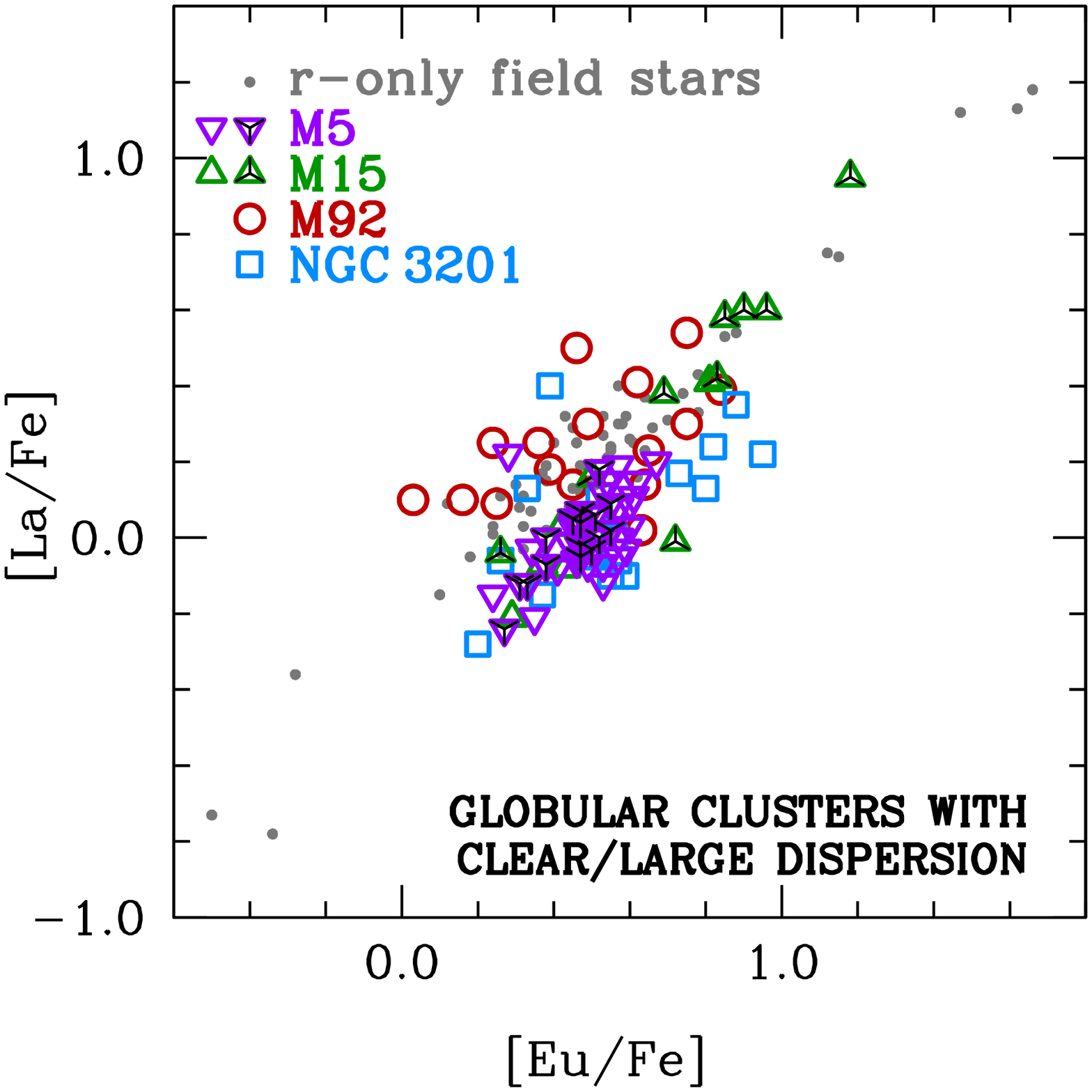} \hspace*{0.1in}
\includegraphics[angle=0,width=2.8in]{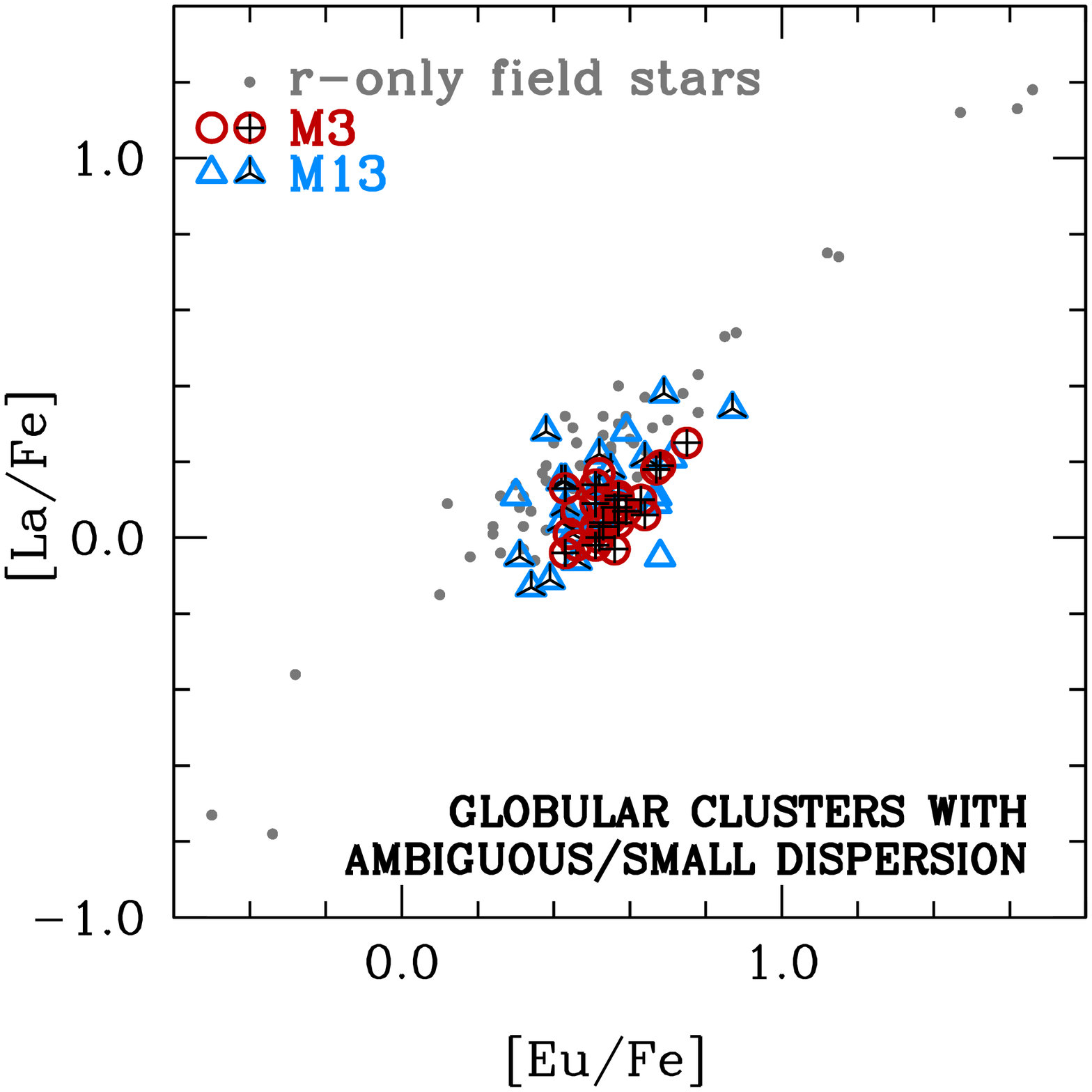} \\
\vspace*{0.10in}
\includegraphics[angle=0,width=2.8in]{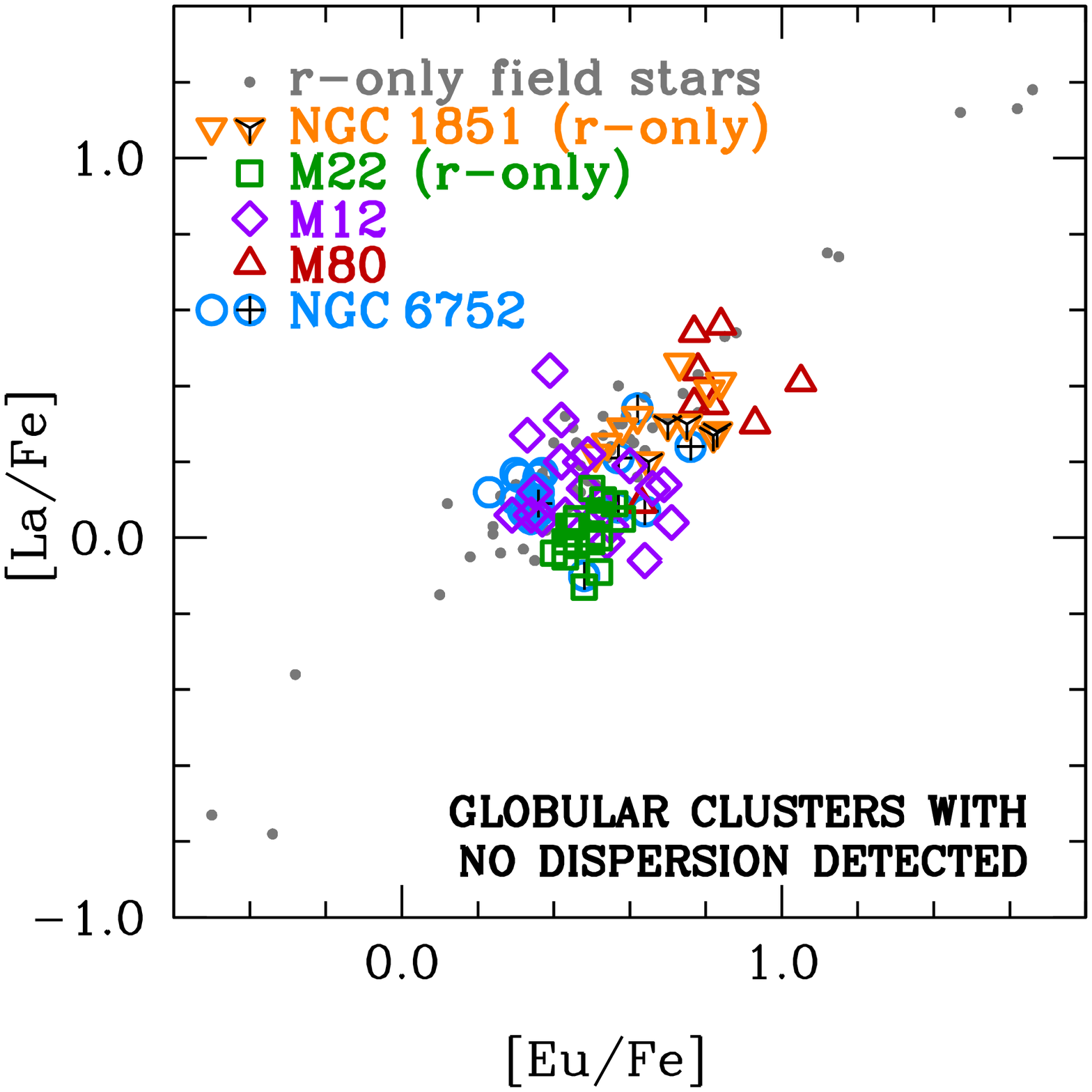} \hspace*{0.1in}
\includegraphics[angle=0,width=2.8in]{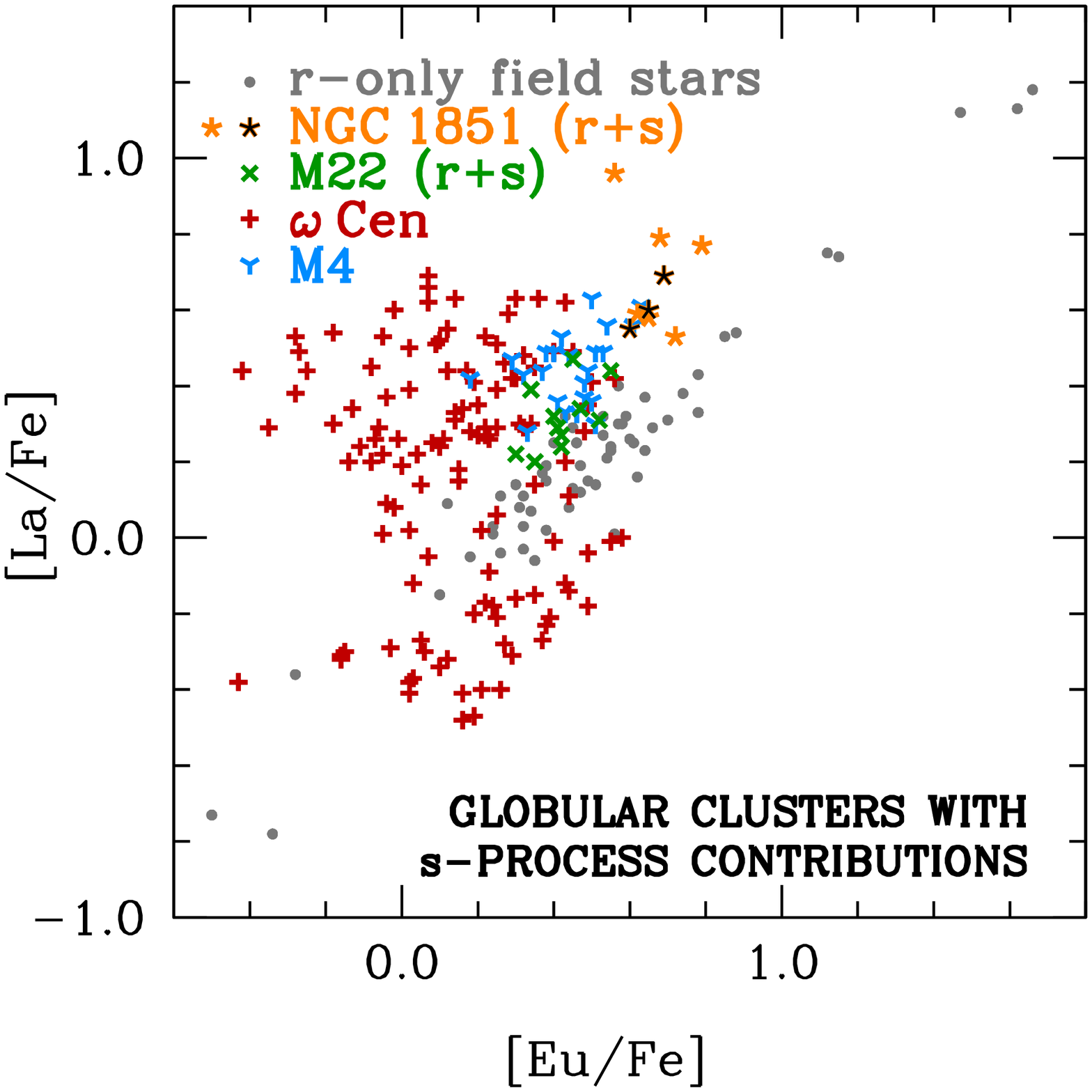}
\end{center}
\caption{
\label{strongplot}
The [La/Fe] and [Eu/Fe] ratios in GCs that exhibit clear
evidence for \rpro\ dispersion (\textit{top left}),
those that exhibit ambiguous evidence (\textit{top right}),
those where no \rpro\ dispersion has been detected (\textit{bottom left}), 
and those with clear evidence for \spro\ enrichment (\textit{bottom right}).
Data are taken from the following sources:
M3 and M13, \citet{sneden04} (crossed) and \citet{cohen05};
M4, \citet{ivans99};
M5, \citet{ivans01} and \citet{lai11} (crossed);
M12, \citet{johnson06};
M15, \citet{otsuki06} and \citet{sobeck11} (crossed);
M22, \citet{marino11};
M80, \citet{cavallo04};
M92, \citet{roederer11}; 
\mbox{NGC~1851}, \citet{yong08a} (crossed) and \citet{carretta10c};
\mbox{NGC~3201}, \citet{gonzalez98};
\mbox{NGC~6752}, \citet{cavallo04} (crossed) and \citet{yong05};
\mbox{$\omega$~Cen}, \citet{johnson10};
\rpro\ field stars, \citet{roederer10b}.
Stars that are common to multiple studies of the same GC
have only been displayed once.}
\end{figure*}

\textbf{\mbox{NGC~3201}}: 
The correlation between [La/Fe] and [Eu/Fe]
is significant. 
\citet{gonzalez98} found a range of nearly 0.7~dex in [Eu/Fe] from
17~stars, nearly 0.7~dex in [La/Fe],
nearly 0.5~dex in [Ba/Fe], and more than 0.4~dex in [Ce/Fe].
The correlation between [Eu/Fe] and [Ba/Fe] is also significant, and
the correlation between [La/Fe] and [Ba/Fe] is suggestive but not
statistically significant.
The large metallicity spread found by Gonzalez \& Wallerstein
has not been reproduced \citep{covey03,carretta09},
but the explanations for the discrepancy suggested by Covey et al.\
should not significantly affect 
La~\textsc{ii} or Eu~\textsc{ii}.
These data suggest that an \rpro\ dispersion is present in \mbox{NGC~3201}.

Abundances in these four GCs are shown in the 
top left panel of Figure~\ref{strongplot}.
A set of metal-poor field stars that exhibit no evidence of
\spro\ enrichment by virtue of their low [Pb/Eu] ratios are
shown for comparison.
The GC data overlay the same region of the diagram
as the field stars.
[La/Eu] is constant in 
these GC stars as both elements vary together, 
and the ratio, [La/Eu]~$= -$0.4~$\pm$~0.2, is consistent
with that observed in metal-poor field stars with 0.0~$<$~[Eu/Fe]~$< +$1.0.
The larger uncertainties in M92 are evident, 
and there may be 1 or 2 stars in M15 and \mbox{NGC~3201} 
that deviate from the main concentration.
These GCs show clear evidence for star-to-star \rpro\ dispersion
relative to Fe.

\textbf{M3} and \textbf{M13}:
\citet{sneden04} and \citet{cohen05} studied a number of 
stars in this pair of GCs, shown
in the upper right panel of Figure~\ref{strongplot}.
The [La/Fe] and [Eu/Fe] ratios in these two studies give conflicting results.
Correlations derived from the Sneden et al.\ data
are highly significant, but the Cohen and Mel{\'e}ndez data
do not show any significant correlation;
statistics for the combined samples produce a highly
significant result for M3 but not M13.
The Sneden et al.\ samples are larger for both GCs.
The ranges spanned by [Ba/Fe] and [La/Fe] are similar in 
both investigations.
Excluding the anomalous star B4.4 in M3, [Eu/Fe] spans 0.3~dex in M3
in the 22~stars of Sneden et al.\ but only 0.1~dex in the 8~stars
of Cohen and Mel{\'e}ndez.
A spurious correlation could arise from poor estimates of the 
stellar parameters; if so, 
we would expect correlations between pairs of other singly-ionized elements
such as
Sc~\textsc{ii} and
Ti~\textsc{ii}.
There is no such correlation 
[($r$, $N$, $P_{c}$)~$=$~($-$0.03, 20, 0.89)]
in the M3 data of Sneden et al., who did not publish
Sc~\textsc{ii} or Ti~\textsc{ii} abundances for M13.
We attribute the discrepancy 
to the different sample sizes and 
conclude that M3 and M13 probably exhibit \rpro\ dispersion.

Three other GCs in our sample, M12, M80, and \mbox{NGC~6752}, 
show no significant correlation between [La/Fe] and [Eu/Fe].
The abundances of M22 and \mbox{NGC~1851}
show a bimodal distribution
\citep{yong08a,marino09,marino11,carretta10c},
which may suggest that they formed under unusual circumstances
(e.g., the merger of two separate GCs).
In M22, the separation is clear in, e.g., [La/Fe] and [La/Eu]
\citep{marino11}. 
The separation in \mbox{NGC~1851} is
less obvious, and we make an approximate division 
at [La/Fe]$_{\rm r} <$~0.50~dex. 
Based on the available observations---illustrated 
in the lower left panel of 
Figure~\ref{strongplot}---M12, M80, \mbox{NGC~6752} and 
the \rpro-only population in M22 
lack \rpro\ dispersion.
The \rpro-only population in \mbox{NGC~1851} shows
a correlation between [La/Fe] and [Eu/Fe] based on 7~stars
from \citet{carretta10c},
while the 5~stars of \citet{yong08a} show no such correlation;
combining these samples produces an ambiguous result
[($r$, $N$, $P_{c}$)~$=$~(0.52, 12, 0.084)].
Additional data may help resolve the matter. 

The lower right panel of Figure~\ref{strongplot} 
demonstrates that \rpro\ dispersion is clearly distinguishable
from \spro\ enrichment.
Stars in M4 \citep{ivans99,yong08b} 
and $\omega$~Cen (e.g., \citealt{smith00})
are enhanced in
\spro\ material, 
and this increases [La/Eu] substantially.
The same effect is observed in the $r+s$ populations
of M22 and \mbox{NGC~1851}.
Except for a few stars in $\omega$~Cen
that exhibit an \rpro-only signature,
the rest are clearly distinct from the
\rpro-only field stars.
The dispersion in
M3, M5, M13, M15, M92, and \mbox{NGC~3201} is unrelated
to \spro\ enrichment.

Finally, we examine whether \rpro\ dispersion correlates with
the classical light element dispersion.
Using [Na/Fe] as a proxy for the light elements, 
we search for correlations between [Na/Fe] and
either [La/Fe] or [Eu/Fe].
As Table~\ref{correlatetab} shows,
there is no correlation in nearly all cases.
In \mbox{NGC~6752} a significant correlation is present in the 6 and 8~stars
of \citet{cavallo04}, but employing the
17 and 38~stars of \citet{yong05} reveals no correlation.
[La/Fe] and [Na/Fe] are anti-correlated 
in the M13 data of \citet{sneden04}, but
this is the opposite of what would be expected 
if Na production is accompanied
by \spro\ nucleosynthesis.
[Eu/Fe] and [Na/Fe] show no significant correlation in M13, so
we conclude that the [La/Fe] versus [Na/Fe] anti-correlation
is not real.
In these 11~GCs, the \rpro\ dispersion is independent of
the light element dispersion.

\section{Discussion}
\label{discussion}

All of these GCs are metal-poor (by selection) and older
than $\approx$~11~Gyr.
We consider whether the GCs that do show \rpro\ dispersion
and those that do not have differences in their 
structural \citep{harris96} or kinematic 
\citep{dinescu99,cassettidinescu07} properties,
horizontal branch morphologies \citep{lee90}, or RR~Lyr types
\citep{clement01}.
We find no differences except that the
GCs with \rpro\ dispersion---including M3 and M13---have 
larger apogalactic radii (10--35~kpc)
than those that do not (3--6~kpc).
(M22 and \mbox{NGC~1851} have large apogalactic radii---9.3 and 
30~kpc, respectively---but as noted previously
their formation histories may be more complex.)
This may indicate that the distribution of \rpro\ material
reflects properties (e.g., mass, formation locations or timescales) 
of the giant molecular clouds (GMCs) in which
the GCs formed (cf., e.g., \citealt{carretta10a}).

\begin{figure}
\includegraphics[angle=0,width=3.4in]{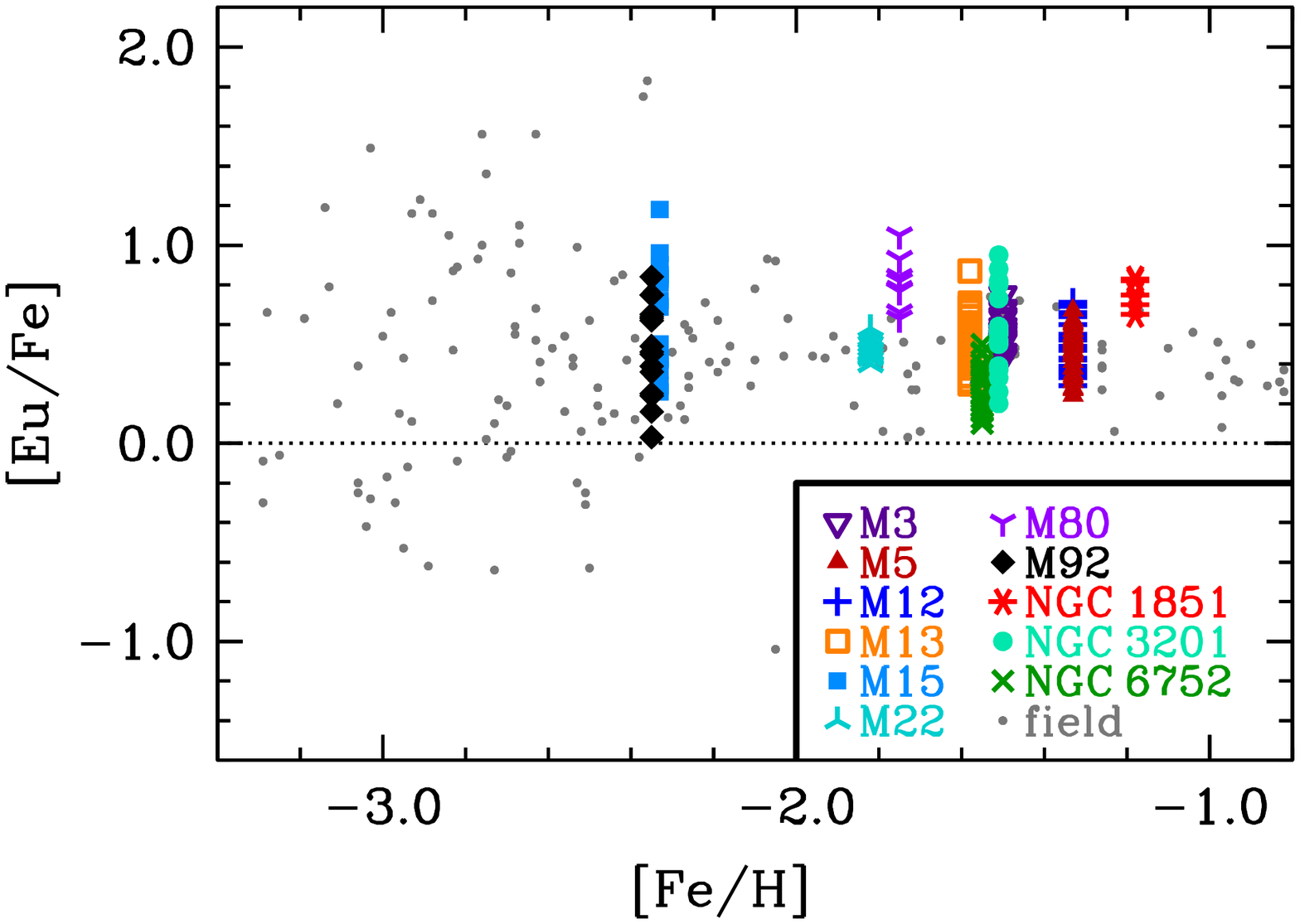}
\caption{
\label{eufeplot}
Comparison of [Eu/Fe] between metal-poor field stars 
and the 11~GCs discussed in Section~\ref{rpro}.
All GC abundances have been compressed to each GC's mean metallicity
(on the \citealt{carretta09} scale).
Just the \rpro-only stars in M22 and \mbox{NGC~1851} have been displayed here.
Field star abundances are taken from the recent literature
\citep{fulbright00,johnson02,honda04,barklem05,francois07,lai08,roederer10a}.
}
\end{figure}

\citet{dorazi10} presented [Ba/Fe] dispersion (root mean squared, rms) 
values for 15~GCs,
including 4 in our sample.
M5, M15, and \mbox{NGC~3201} have [Ba/Fe] rms values of 0.210, 0.412, and
0.267, respectively, while \mbox{NGC~6752} has an rms of 0.176.
This suggests that $\approx$~0.2 may represent the 
detectable \rpro\ dispersion threshold.
Based on this simple-minded discriminant 
 (and ignoring GCs with [Fe/H]~$> -1.0$ or [Ba/Fe]~$> +0.4$,
which may contain significant \spro\ material),
from the D'Orazi et al.\ data we predict that 
M10, 
M55, 
M68, 
and \mbox{NGC~6397} may
exhibit \rpro\ dispersion, while 
M30 
and 
M79 
may not.

As has been discussed in detail elsewhere 
(e.g., \citealt{gratton04}),
the heavy element abundances in metal-poor GCs and field
stars are very similar.\footnote{
Figure~12 of \citet{roederer10b} indicates that
GCs have ratios of light to heavy \ncap\ elements,
e.g., [Y/Eu], similar to those found in field stars.}
We illustrate this fact in Figure~\ref{eufeplot}
for the 11~GCs analyzed in Section~\ref{rpro}.
The mean [Eu/Fe] for each of these GCs is $+$0.5~$\pm$~0.3,
which is consistent with the majority of field stars with 
$-$2.5~$<$~[Fe/H]~$< -$1.0.
The presence of \rpro\ dispersion in some
GCs is not inconsistent with the idea that
GCs were once much more massive
(e.g., \citealt{dercole08,carretta10a,gratton10})
and lost a significant fraction of their first generation stars
(i.e., those with undepleted O and unenhanced Na)
to the stellar halo (cf.\ \citealt{martell10}).

Nearly all metal-poor field stars contain detectable
amounts of \rpro\ material,
and all GC stars studied to date
have detections or uninteresting upper limits for these elements.
\citet{roederer10b} uncovered
a range of [La/Eu] ratios (spanning at least 0.5~dex)
in field stars with low [Pb/Eu] (i.e., lacking \spro\ material).
This suggests that the variations are intrinsic to the \rpro\ yields, since
dilution of material from identical but rare \rpro\ events cannot alone 
account for a range of [La/Eu].
In GCs, homogeneous Fe-group abundances and
inhomogeneous \rpro\ abundances relative to Fe could indicate
that \rpro\ material is not produced by every SN 
or is unevenly dispersed 
(e.g., concentrated in jets; \citealt{otsuki06}).
Alternatively, \rpro\ material may have been injected into 
the gas shortly before star formation, thereby limiting 
its homogenization (\citealt{sneden97}; Otsuki et al.).
Either way, \rpro\ material must be present but
incompletely mixed into the gas from which some GCs form.

Expectations from O/Ne/Mg-core SN physics 
(e.g., \citealt{wheeler98,wanajo03})
predict that low-mass ($\sim$~8--10~\msun)
Type~II SNe are a dominant source of \rpro\ material
but produce little or no Fe-group material. 
If so, they would be among the last sources contributing to the 
chemical inventory of the GC ISM before massive AGB stars undergo
significant mass loss.
Yet, some (unidentified) mechanism(s) 
must exist to prevent \rpro\ material 
from fully homogenizing
until other sources have polluted the GC ISM.
Otherwise we would observe \rpro\ dispersion only
in first generation stars with undepleted O and unenhanced Na;
later generations would have homogeneous \rpro\ abundances,
which is not observed \citep{smith08}.
If the dynamical crossing time of the GMCs
from which GCs form is comparable to the mixing timescale
(e.g., \citealt{mckee07} and references therein; see also
\citealt{carretta10b}), we can estimate a lower limit 
for the timescale of chemical homogenization.
8--10~\msun\ stars have lifetimes of a few ($\lesssim$~10) Myr,
comparable to 
the crossing times
($\sim$~5--20~Myr) of 10$^{5}$--10$^{7}$~\msun\ GMCs.
If unmixed \rpro\ material is preserved for several tens of Myr 
(e.g., in the atmospheres of main sequence stars, \citealt{gratton10})
before being returned into the ISM,
it could later become diluted with
the ejecta of high-mass AGB stars that may produce the
familiar O-Na and Mg-Al anti-correlations in subsequent stellar generations.

Present-day GC masses are large enough to ensure 
complete sampling (on average) of a Salpeter IMF
($M > 10^{5} M_{\odot}$),
and their initial masses would certainly have been so,
although star formation in proto-GCs clearly extended longer than
a single burst.
This may help explain why the mean ratios of elements
produced in Type~II SNe are generally constant from one GC to the next.
(Stochastically sampling the IMF could significantly affect 
the abundances in lower-mass systems like the
ultra-faint dwarf galaxies; e.g., \citealt{koch08}, 
\citealt{feltzing09}, \citealt{simon10}.)
Adopting the range of \rpro\ yields predicted by the
high entropy neutrino wind simulations of \citet{farouqi10}
(10$^{-6}$ to 10$^{-4}$~\msun\ per event), assuming
$M_{\rm r,~total} \sim 10^{4} \times M_{\rm Eu}$,
and assuming a star formation
efficiency of 10\% in GCs with initial masses $\sim$~10 times larger
than their present mass 
predicts a lower limit of $\sim$~10$^{2}$--10$^{4}$ SNe
to produce the observed GC \rpro\ enrichment levels.
Alternatively we can use the observed GC \rpro\ abundances 
to derive a lower limit on the
fraction of 8--10~\msun\ SNe 
that produce an \rpro.
Assuming the maximum \rpro\ yields
from Farouqi et al.\ and assuming that all \rpro\ material is 
incorporated into stars, 
we derive a lower limit of $\sim$~1/7 SNe with initial masses
8--10~\msun\ should produce an \rpro.
Relaxing these assumptions would increase this fraction substantially. 

The data currently available indicate that \rpro\
dispersion may be a common but not ubiquitous
characteristic of metal-poor GCs.
These data are inadequate to discern details such as 
the fraction of GCs with dispersion or the range
and distribution of \rpro\ abundances within each GC.
Whatever the explanation,
an understanding of what parameter(s) control the distribution of
\rpro\ material in GCs is greatly desired.

\acknowledgments

We thank A.\ McWilliam for insightful discussions,
E.\ Carretta, A.\ Marino, and J.\ Sobeck for sharing their 
results in advance of publication, and the referee
for providing a positive and useful report.
I.U.R.\ is supported by the Carnegie Institution of Washington 
through the Carnegie Observatories Fellowship.


\end{document}